\documentclass[prd,aps]{revtex4}
\usepackage{amssymb,graphicx}

\newcommand{\bm}[1]{\mbox{\boldmath$#1$}}
\newcommand{\Real}{\mathbb{R}}

\begin{document}
\renewcommand{\arraystretch}{2.}
\setlength{\tabcolsep}{3mm}

\title{Summation by parts and dissipation for domains with excised regions}
\author{Gioel Calabrese$^{1,2}$, Luis Lehner$^1$, Oscar Reula$^3$, 
Olivier Sarbach$^1$, and Manuel Tiglio$^{1,4,5}$}
\affiliation{$1$ Department of Physics and Astronomy, Louisiana State
University, 202 Nicholson Hall, Baton Rouge, LA 70803-4001\\
$2$ School of Mathematics, University of Southampton, Southampton, SO17 1BJ, UK
\\
$3$ FaMAF, Universidad Nacional de Cordoba, Cordoba, Argentina 5000\\
$4$ Center for Computation and Technology, 302 Johnston Hall, Louisiana State
University, Baton Rouge, LA 70803-4001 \\
$5$ Center for Radiophysics and Space Research, Cornell University, Ithaca, NY 14853}

\begin{abstract}
We discuss finite difference techniques for hyperbolic equations in
non-trivial domains, as those that arise when simulating black hole
spacetimes. In particular, we construct dissipative and difference
operators that satisfy the {\it summation by parts} property in domains
with excised multiple cubic regions. This property can be used to
derive semi-discrete energy estimates for the associated
initial-boundary value problem which in turn can be used to prove
numerical stability.
\end{abstract}

\maketitle

\section{Introduction}

An important problem in astrophysics is to model in a detailed way the
collision of two black holes \cite{CutlerThorne,Hughes}. This requires
numerically integrating Einstein's field equations and extracting from
the simulations the relevant physical information.  Unfortunately, it
is difficult to obtain numerical solutions of these equations in
generic three-dimensional settings, especially for long term
simulations. Obstacles to this goal are encountered in both the
analytical and numerical arenas. In the {\em analytical} one, the
formulation of a well posed initial-boundary value problem is not completely
understood. This includes the definition of proper initial and
boundary conditions and the equations determining the future evolution
of the fields. In the {\em numerical arena} one seeks to define
numerical techniques that allow for long term accurate
evolutions. This requires the construction of appropriate discrete
operators to implement the initial-boundary value problem. To date,
despite considerable advances in both fronts\footnote{For some reviews
on the subject see~\cite{reviews} and references cited therein, and
for a description of more recent efforts see~\cite{3Dexcision}}, the
challenge of simulating generic three-dimensional black hole systems
remains unattained.

This article intends to provide some initial steps for setting up
numerical techniques suitable to address the numerical stability of
equations like the ones in question, by extending and devising finite
difference techniques to tackle first order symmetric hyperbolic
problems in non-trivial domains, with numerical stability being
guaranteed in the linear case. Furthermore, via a local argument, one
can assert that these methods should be useful in evolving smooth
solutions of quasi-linear symmetric hyperbolic equations as well, as
is the case of the full, non-linear Einstein vacuum equations when
appropriately written \cite{reula}. Although the main motivation of
this work is to present techniques for the simulation of Einstein's
equations on domains with excised regions, the techniques here
presented are readily applicable to any symmetric hyperbolic problem
in such domains. Applications of these techniques in a variety of
scenarios will be presented elsewhere
\cite{lsulong2,davegioel,bubble1,olivieraxis,hyperGR}.

This work is organized in the following way. In sections
\ref{Sect-NumStab} and \ref{Sect-OB} we review some of the issues
involved in obtaining stable numerical schemes through the energy
method, to set the stage for the specialized discussions that
follow. In section \ref{Sect-excision}, the main new results of the
article are presented: we derive three dimensional difference
operators satisfying summation by parts for non-trivial domains which
enable one to obtain energy estimates. We further introduce

dissipative operators which do not spoil these estimates. To complete
a stability proof, one needs to impose boundary conditions without
affecting the semi-discrete energy estimates obtained thanks to the
use of the constructed dissipative and difference operators. One way
of doing so is by imposing boundary conditions through an orthogonal
projection, as done by Olsson \cite{olsson}, a technique which can
even be applied to non-smooth domains. Conclusions are drawn in
section \ref{Sect-Conclusions}. The appendices include technicalities,
discussions about Courant limits and further details about the
implementation of boundary conditions.

\section{Numerical stability through the energy method: an overview}
\label{Sect-NumStab}

In order to construct stable finite difference schemes for
initial-boundary value problems (IBVPs) associated with partial
differential evolution equations we will use the method of lines
\cite{GKO}. This means that we first discretize the spatial
derivatives appearing in the partial differential equations so as to
obtain a large system of ordinary differential equations for the grid
functions. This system is usually called {\it semi-discrete
system}. In the following, we assume that the solutions of the partial
differential equations satisfy an {\it energy estimate} which bounds
some norm of the solution at some fixed time $t$ in terms of a
constant $C = C(t)$ which is independent of the the initial and
boundary data times the same norm of the initial data and a bound on
the boundary data. As we will discuss in more details below, one can
derive a similar energy estimate for the semi-discrete system if the
partial derivatives and the boundary conditions are discretized in an
appropriate way. Provided that the constant $\tilde{C}$ involved in
this semi-discrete estimate can be chosen to be resolution-independent
for small enough resolution, this immediately implies numerical
stability for the semi-discrete problem. An important ingredient in
the derivation of such estimates is the {\it summation by parts} (SBP)
property which is the discrete analogue of {\it integration by parts}.

There are many possible discretizations of the partial derivatives and
the boundary conditions which are consistent with the IBVP and which
yield a stable semi-discrete scheme, and in general, the constant
$\tilde{C}$ will be larger than the corresponding constant $C$ of the
continuum problem. By carefully choosing the discretization one can
achieve optimal semi-discrete energy bounds, in the sense that
$\tilde{C}=C$ which means that the norm of the solution to the
semi-discrete problem satisfies the same estimate as the norm of the
solution to the analytic solution. Alternatively or complementary, one
might want to add artificial dissipation in order to control the high
frequency modes of the solution which are always poorly represented
with finite difference approximations.

Finally, by discretizing the time derivatives one obtains the fully
discrete system which is numerically implemented. If the semi-discrete
system is stable one can show that the fully discrete system is stable
as well, provided an appropriate time integrator is chosen. This will
be briefly discussed later on.

To summarize, numerical stability in this approach reduces to
\begin{itemize}
\item Formulate a well posed IBVP for the
  problem one wants to solve at the continuum. 
\item Construct difference operators for the domain of interest that
  satisfy the SBP property. 
\item Optionally or even complementary:
\begin{itemize} 
\item Construct dissipative operators that do not spoil the energy estimate.
\item Rearrange the semi-discrete
  equations to achieve optimal energy estimates.  
\end{itemize}
\item Impose boundary conditions without spoiling the semi-discrete stability.
\item Achieve fully discrete stability by appropriately 
choosing the time integrator.
\end{itemize}
The next subsections give an overview of these points. Section
\ref{Sect-OB} discusses some issues that appear in the IBVP case,
while section \ref{Sect-excision} presents results for the case in
which there are inner boundaries and excised regions in the
computational domain.

\subsection{Semi-discrete stability}

\subsubsection{Well posedness}
\label{WP}
There are many definitions of well posedness though roughly, and
without going into too many technical details, an IBVP is said to be
well posed if
\begin{enumerate}
\item a local in time solution with certain smoothness exists, 
\item the solution is unique, and, 
\item the solution depends continuously on the initial and boundary
data of the problem.
\end{enumerate}
There are different approaches to obtaining well posed formulations of
a given problem. A common one is the so called energy method
\cite{GKO} where one seeks an energy norm which has the form
\begin{equation}
{\mathcal E}(t) = ||u(t,.)||^2 = (u,u),
\label{energynorm}
\end{equation}
where $u(t,.)$, the solution of the IBVP at some given time $t$, lies
in some Hilbert space with scalar product $(.,.)$.  In many physical
situations, this Hilbert space can be motivated by the existence of a
conserved energy. In general however, the expression ${\mathcal E}$
does not need to coincide with any physical energy.  For first order
symmetric hyperbolic linear systems, the Hilbert space can be taken to
be the space of square integrable functions over some domain $\Omega$
with scalar product
\begin{equation}
(u,v)  = \int_{\Omega} u \cdot H  v \, dx  \;,
\label{scalar_product}
\end{equation}
where the symmetrizer $H$ is a symmetric positive definite
matrix-valued function on $\Omega$ which is bounded from above and
away from zero, and ${u} \cdot {v}$ denotes the standard scalar
product between the vector-valued functions $u$ and $v$. In this case,
and for suitable boundary conditions, one has an a priori estimate of
the form
\begin{equation}
{\mathcal E}(t) \leq C(t) {\mathcal E}(0) + G(t), \quad t \geq 0,
\label{energygrow}
\end{equation}
where $C(t)$ is independent of the initial and boundary data, and
$G(t)$ bounds the boundary data. This means that the solution at time
$t$ can be bounded by the energy at $t=0$ and a bound on the
energy pumped in through the boundary of the domain. In
particular, for each fixed $t \geq 0$, small variations in the data
result in small variations of the solution.

As an example, consider the initial-boundary value problem
\begin{eqnarray}
&&\partial_t u(t,x) = -\partial_x u(t,x),
\qquad x > 0,\quad t > 0,
\label{Eq:Ev}\\
&&u(0,x) = f(x), \\
&&u(t,0) = g(t),
\label{Eq:BC}
\end{eqnarray}
with smooth initial and boundary data, $f(x)$ and $g(x)$, respectively,
that satisfy appropriate compatibility conditions at $(t,x) = (0,0)$.
We also assume appropriate fall-off conditions at $x = \infty$. 
Taking a time derivative of the energy (\ref{energynorm}) with $H = 1$
and using the evolution equation (\ref{Eq:Ev}) we obtain
\begin{equation}
\frac{d}{dt} {\mathcal E} = -(u,\partial_x u) - (\partial_x u,u) = u^2(t,0), 
\end{equation}
where we have used integration by parts in the last step.
Using the boundary condition (\ref{Eq:BC}) and integrating
one obtains the energy estimate (\ref{energygrow}) with
$C(t) = 1$ and $G(t) = \int_0^t g(\tau)^2 d\tau$.

The energy method is not only a valuable tool in studying the system
at the analytical level, but it can also be used to produce stable
numerical discretizations by considering a discrete analogue.
Essentially, this involves defining discrete operators with
which one reproduces, at the discrete level, the steps taken at the
continuum when obtaining the energy estimate. In what follows we
briefly review the steps involved.

Before we proceed, we mention that the existence of an a priori energy
estimate, where the energy norm has the form of an integral over a
local density expression, is a sufficient but not necessary condition
for well posedness \cite{GKO}. A different approach for analyzing
stability is the Laplace method \cite{Laplace} which gives necessary
conditions for well posedness. However, the application of this method
to obtain sufficient conditions is rather cumbersome and problem
dependent. On the other hand, the energy method is considerably
simpler and so it is preferred when applicable.  This is the approach
we shall follow in this article.  It can be applied to a very large
set of physical problems.

\subsubsection{Summation by parts}
\label{SBP}
As we have seen, a tool that is used in the derivation of energy
estimates is integration by parts. In one dimension (1D) it reads
\begin{displaymath}
\int_a^b v(x)\cdot \partial_x u(x)\,dx  
 + \int_a^b \partial_x v(x) \cdot u(x)\,dx
 = \left. v(x) \cdot u(x)\right|_a^b \; ,
\end{displaymath}
where $a < b$. Let $N$ be a positive integer, $\Delta x = (b-a)/N$ and
consider the grid defined by $x_j = a + j\Delta x$, $j=0,...,N$. The
function $u(x)$ is approximated by a grid function $(u_j) =
(u_0,...,u_N)$. One of the ingredients needed to obtain an energy
estimate for the semi-discrete problem is to construct difference
operators D (approximating the first order partial differential
operator $\partial/\partial x$) which satisfy a SBP rule \cite{GKO}
\begin{equation} (v,Du)_{\Sigma }^{[0,N]} + (Dv,u)_{\Sigma }^{[0,N]}=
\left. v_j \cdot u_j\right|_{j=0}^N \;, 
\label{eq:sumbyparts}
\end{equation} 
with respect to some scalar product
\begin{equation}
(u,v)_{\Sigma }^{[0,N]} = \Delta x\sum_{i,j=0}^N \tilde{\sigma}_{ij} u_i \cdot v_j \, .
\label{scalar_product2} 
\end{equation}
Here, the weights $\Sigma \equiv (\tilde{\sigma}_{ij})$ must be chosen
such that the norm $E = \|u\|_{\Sigma }^2 = (u,u)_{\Sigma }$, which
can be seen as the discrete version of that defined in
Eq. (\ref{energynorm}), is positive definite. Achievement of SBP in
general requires a careful choice of {\em both} the scalar product and
difference operator; for a given choice of scalar product there might
not exist any difference operators satisfying the SBP property.  In
the following, we skip the label $\Sigma$ when it refers to the
trivial scalar product $\tilde{\sigma}_{ij} = \delta_{ij}$. Also, when
there is no need to specify the domain, we skip the superscript
$[0,N]$.

Having a 1D operator that satisfies SBP with respect to a {\em
diagonal scalar product} $\tilde{\Sigma } =
(\tilde{\sigma}_{ij})=(\delta_{ij}\sigma_{i})$, one can construct a 3D
operator by simply applying the 1D difference operator in each
direction. The resulting 3D operator satisfies SBP with respect to the
scalar product
\begin{displaymath}
(u,v)_{\Sigma } = \Delta x \Delta y \Delta z \sum_{ijk} \sigma_{ijk}
u_{ijk}\cdot v_{ijk}\;, 
\end{displaymath}
with coefficients (making a slight abuse of notation)
$\sigma_{ijk}=\sigma_{i} \sigma_{j} \sigma_{k} $ \cite{olsson}. This
is {\em not} necessarily true if $\tilde{\Sigma}$ is not diagonal. For
this reason, we will only consider diagonal scalar products in this
article. In the absence of boundaries, SBP reduces to
\begin{equation}
(v,Du)_{\Sigma}^{(-\infty,\infty)} + (Dv,u)_{\Sigma}^{(-\infty,\infty)} = 0,
\label{isbp}
\end{equation}
provided suitable fall off conditions are imposed. In this case 
simple 1D operators that satisfy SBP with respect to the trivial
scalar product are the standard centered difference operators $D=D_0$; for example,
\begin{eqnarray}
& (D_0 u)_j = \frac{u_{j+1} - u_{j-1}}{2\Delta x}  & \hbox{(second order accurate)}
\label{Eq:SecondSD}\\
& (D_0 u)_j = \frac{-u_{j+2} + 8 u_{j+1} - 8 u_{j-1} + u_{j-2}}{12\Delta x} & 
\hbox{(fourth order accurate)}
\label{Eq:FourthSD}
\end{eqnarray}
etc. Defining a 3D difference operator by just applying this
one-dimensional one in each direction will satisfy SBP with respect to
the trivial 3D scalar product.

As a simple application, consider the initial value problem for the
symmetric hyperbolic system,
\begin{eqnarray}
&& \partial_t u(t,x) = A^i\partial_i u(t,x) + B u(t,x),
\qquad x \in \Real^3, t > 0,
\label{Eq:HypEq}\\
&& u(0,x) = f(x),
\label{Eq:ID}
\end{eqnarray}
where $u(t,x)$ is a vector valued function, the initial data $f(x)$
vanishes for $|x|$ sufficiently large, and where the matrices $A^i$
are symmetric. Assume for simplicity that the $A^i$ are constant and
that $B=0$. At the analytic level, the energy norm defined by
Eq. (\ref{energynorm}) with $H=1$ satisfies ${\mathcal E}(t) =
{\mathcal E}(0)$. If we replace $\partial_i$ by the centered
differencing operator $D_i$ in the $i$'th direction, the discrete
energy $E(t) = \| u(t) \|^2$ with respect to the trivial discrete
scalar product satisfies the same estimate: $E(t) = E(0)$, and the scheme is
stable by construction.

\subsubsection{Dissipation }
\label{DISIP}

Even though one can achieve numerical stability through SBP, it is
sometimes convenient to add artificial dissipation to the problem. One
possible reason for doing so is the presence of high frequency modes
which -- even if they go away with increasing resolution -- grow in
time at fixed resolution. Although this would not be the case in our
present discussion, it is well known that the addition of dissipation
can aid in stabilizing schemes that would otherwise be numerically
unstable. An example of this would be the case of system that is stable 
at the semidiscrete level --for example, a symmetric system with discretized
with operators satisfying SBP-- but becomes unstable because the region of 
absolute stability of the time integrator does not contain a neighborhood 
of the origin along the imaginary axis.

As an example, consider the initial value problem
(\ref{Eq:HypEq},\ref{Eq:ID}).  The standard way \cite{GKO} to
introduce dissipation in the problem is to modify the right-hand side
(RHS) of the equations
\begin{displaymath}
A^i \partial_i u + B u \to A^i D_i u + Bu + Q_{d}^{(x)} u + Q_{d}^{(y)}
u + Q_{d}^{(z)} u,
\end{displaymath}
where $D_i$ is a differential operator satisfying SBP and where $Q_d$
is a differential operator that vanishes in the limit of infinite
resolution, such that the consistency of the scheme is not altered,
and such that
\begin{equation}
(u,Q_du)_{\Sigma } \leq 0 \qquad
\hbox{for all $u$}\; ,
\label{dissipation}
\end{equation} 
with some inner product $\Sigma $ for which a discrete energy estimate
holds (for example, the one with respect to which SBP holds). The
dissipative property (\ref{dissipation}) ensures that the discrete
energy estimate is not spoiled and might be useful to stabilize an
otherwise unstable scheme. Furthermore, $Q_d$ can be chosen such that
it controls spurious high frequency modes in the solution.
 
As mentioned above, in the absence of boundaries, standard centered
derivatives satisfy SBP with respect to the trivial scalar
product. Similarly, the Kreiss--Oliger dissipation \cite{ko}
\begin{equation}
Q_d = \sigma (-1)^{r-1} (\Delta x)^{2r-1} D_+^r D_-^r \, ,
\qquad \sigma \geq 0,
\label{KO}
\end{equation}
where
\begin{equation}
D_+ u_j = \frac{u_{j+1}-u_j}{\Delta x} \; ,
\qquad
D_- u_j = \frac{u_j - u_{j-1}}{\Delta x} \;, 
\label{Eq:OneSided}
\end{equation}
denote the one-sided difference operators, satisfies the dissipative
property (DP) with respect to that scalar product. Furthermore, if the
accuracy of the scheme without artificial dissipation is $q$, choosing
$2r-1 \geq q$ does not affect the accuracy of the scheme. Notice also
that with a slight abuse of notation we have used $\sigma$ to denote
the dissipative parameter; this should not cause confusion with
respect to $\sigma_{i} $ used in the context of weights of scalar
products, since the latter have subindices.

\subsubsection{Optimal energy bounds}
\label{stricstability}

So far, we have discussed how to obtain energy estimates for the
semi-discrete problem by constructing finite difference operators that
satisfy SBP. These estimates imply numerical stability,
by providing a discrete analogue of Eq. (\ref{energygrow}). Although
the resulting scheme is stable the discrete estimate in principle
might not agree with that found at the continuum; that is, the
constant $\tilde{C}(t)$ appearing in the discrete analogue of
Eq. (\ref{energygrow}) might be larger than $C(t)$. If one can show
that $\tilde{C}(t) = C(t)$ we say that the scheme is {\em strictly
stable}\footnote{Note that our definition of strict stability defers
from that given in Ref. \cite{GKO}, where the scheme is called
strictly stable if $\tilde{C}(t) = C(t) + O(\Delta x)$.}. The solution
to non-strictly stable schemes can have unwanted features such as
artificial growth in time of the errors. In the limit $\Delta x
\rightarrow 0$ these features would disappear; but one would like to
avoid or minimize them even at fixed resolution.

For first order symmetric hyperbolic linear systems, like the one
described by Eqs. (\ref{Eq:HypEq},\ref{Eq:ID}), one can achieve strict
stability \cite{olsson} by rewriting the partial differential equation
in skew-symmetric form:
\begin{equation}
\partial_t u  = \frac{1}{2}A^i D_i u + \frac{1}{2} D_i(A^i u)
+\left(B - \frac{1}{2}\partial_i(A^i)\right)u \, .
\label{Eq:SkeySymForm}
\end{equation}
One can show that the resulting scheme is strictly stable with respect
to the energy $E(t) = \| u(t) \|^2$ defined by the trivial discrete
scalar product. In contrast to this, the simple discretization
$\partial_t u = A^i D_i u + B u$, while yielding a stable scheme, does
not necessarily yield a {\em strictly} stable scheme if the matrices
$A^i$ are not constant. Strict stability is particularly useful if the
formulation at the continuum admits a sharp estimate. In this case,
the construction of a strictly stable scheme can be exploited to rule
out artificial growth of the solutions \cite{lsulong2}. Consider, for
example, a system with time-independent coefficients,
\begin{equation}
\partial_t u = A^i(x) \partial_i u + B(x) u \label{vcoeffs} \, .
\end{equation}
Let $H$ be a symmetrizer, i.e. a symmetric positive definite
matrix-valued function $H(x)$ which is bounded from above and away
from zero and which is such that the matrices $H A^i$ are
symmetric. If $H$ can be chosen such that
\begin{equation}
\partial_i(HA^i) = HB + (HB)^T\; ,
\label{nat_sym}
\end{equation}
one can show that the energy given by Eq. (\ref{energynorm}) is
conserved, and that by rewriting the semi-discrete equations as
\begin{displaymath}
\partial_t u  = \frac{1}{2}A^iD_i u + \frac{1}{2} H^{-1} D_i (HA^i u)
+\left(B - \frac{1}{2}H^{-1}\partial_i(HA^i)\right)u \; ,
\end{displaymath}
the semi-discrete energy does not grow either. Therefore the numerical
solution will not have spurious growth even at a fixed resolution. In
principle, there is no reason why such a symmetrizer should
exist. However, on physical grounds, there should be one whenever
there is a well defined local energy density.

The usefulness of having a conserved energy at the semi-discrete level
is discussed in detail in Ref. \cite{lsulong2}.

\subsection{Fully discrete stability}

Proceeding through the steps above described (\ref{WP}-\ref{SBP}), and
optionally (\ref{DISIP}-\ref{stricstability}), one obtains an energy
estimate for the {\em semi-discrete problem} which implies numerical
stability for this semi-discrete system.

However, one is of course interested in the stability of the {\it
fully discrete problem}. One particular simple but powerful approach
to achieve this goal is to follow the strategy based on the method of
lines and employ a time-integrator guaranteed to give rise to a stable
scheme. A useful feature of this approach is that one can derive
conditions for the time integrator that are sufficient for fully
discrete stability and independent of the details of the spatial
discretization \cite{levy_tadmor,tadmor,kreiss_wu}. Two rather
straightforward options are given by third and fourth order
Runge-Kutta schemes \cite{levy_tadmor,tadmor,kreiss_wu}. In appendix
\ref{App-CFL} we present an elementary discussion, for a wave equation
in a domain without boundaries, that gives an idea of possible values
of the Courant factor. For more complicated equations and domains see
Ref. \cite{tadmor}.

So far our discussion has ignored the presence of boundaries. However
these are unavoidable in most problems of interest. In the next
sections we discuss how to modify the previously described techniques
such that one recovers an energy
estimate in the presence of boundaries. For the sake of clarity 
in the presentation, we first
concentrate on simple domains which do not posses inner boundaries and
then consider the case of domains with holes.

\section{Numerical stability for IBVPs}
\label{Sect-OB}

We here discuss appropriate finite-difference schemes for simple 3D
domains. Consider the following IBVP in the cubic domain $\Omega =
[0,1]\times [0,1]\times [0,1]$, with maximally dissipative boundary conditions 
($\Gamma$ denoting its (non-smooth) boundary),
\begin{eqnarray}
\partial_t u(t,x) &=& A^i\partial_i u(t,x) + B u(t,x), 
\qquad x \in \Omega,\, t > 0, 
\label{eq:symhypsystem} 
\\ u(0,x) &=& f(x), \qquad\qquad\qquad\qquad\quad\;\, x\in \Omega,\\ 
w_+ (t,x) &=& S(x) w_-(t,x) + g(t,x), \qquad x \in \Gamma,\, t > 0,
\label{eq:MaxDiss}
\end{eqnarray}
where the matrices $A^i$ are symmetric and $w_{\pm}(t,x)$ represent the
in- and outgoing characteristic variables with respect to the unit
outward normal to the boundary. It is assumed that the coupling matrix
$S(x)$ does not depend on time and is small enough so as to imply an
energy estimate \cite{rs}. It is also assumed that the initial and
boundary data are smooth enough and satisfy compatibility conditions
at the intersection between the initial surface and boundaries.

\subsection{Summation by parts}

If one discretizes the RHS of equation (\ref{eq:symhypsystem}) according to
\begin{equation}
\partial_t u(t) = A^i D_i u(t,x) + B u(t,x), \qquad x \in \Omega,\, t > 0,
\label{eq:symhypsystem_discrete}
\end{equation}
where the discrete derivative operator $D_i$ satisfies SBP, one will
obtain an energy estimate, modulo boundary contributions that appear
after SBP. As already mentioned, a simple strategy to construct 3D
operators satisfying SBP is to use in each direction a 1D operator
satisfying SBP with respect to a diagonal scalar product, and this is
what we do in what follows.

Existence proofs of high order difference operators and scalar
products satisfying SBP in 1D with boundaries can be found in
\cite{Kreiss-Scherer}.  Explicit expressions for these operators can
be found in \cite{strand,GKO} (some of these operators have
non-diagonal associated scalar products, and therefore their use
beyond one dimensional cases does not guarantee numerical
stability). In the interior the derivatives are approximated by one of
the centered finite difference operators mentioned in the previous
section, and the operator is modified near the boundary points. The
simplest example is
\begin{equation}
Du_0 = D_+ u_0\, , \;\;\;  
Du_j= D_0 u_j \; (j=1,\ldots,N-1), \;\;\; 
Du_N = D_- u_N\, , \label{eq:Strand21}
\end{equation}
where the operators $D_0,D_+,D_-$ are defined in
Eqs. (\ref{Eq:SecondSD}) and (\ref{Eq:OneSided}). The operator
(\ref{eq:Strand21}) is second order accurate in the interior and first
order at the boundaries. It satisfies SBP, Eq. (\ref{eq:sumbyparts}),
with respect to the diagonal scalar product $\sigma_0 = \sigma_N =
1/2$, $\sigma_j = 1$ for $j=1,\ldots, N-1$. For high order difference
operators, the operator and the weights in the scalar product may
needed to be modified at points that are near the boundary points as
well.

If the boundary is not smooth and possesses corners and edges the
prescription remains the same, provided the vertices and edges are
{\em convex}. Since in the diagonal case the 3D scalar product is just
the product of the 1D one, the weights $\sigma_{ijk}$ are $1 \times 1
\times 1 = 1$ in the interior points, and in the outer boundary:
$1\times 1 \times 1/2 = 1/2$ at the boundary faces, $1\times 1/2\times
1/2=1/4$ at the edges and $1/2\times 1/2 \times 1/2 = 1/8$ at the
vertices. The presence of {\em concave edges and vertices} requires
modifications, which need to be treated carefully. These will
arise, for instance, when considering a computational domain with an
interior excised region. We discuss this case in section IV.

\subsection{Artificial dissipation}
\label{artdiss}

Having introduced the simple difference operator (\ref{eq:Strand21})
that satisfies SBP in a cubic domain with respect to the 1D scalar
product $\mbox{diag}(1/2,1,1,\ldots,1,1,1/2)$, we now construct some
operators that satisfy the dissipative property with respect to the
same scalar product. We start with the dissipation operator (\ref{KO})
for $r=1$ but redefine $\sigma$ and rewrite the operator as
\begin{equation}
Q_d = \sigma ( \Delta x)^s D_+ D_- \; .
\label{dis2}
\end{equation}
In the absence of boundaries, it satisfies the DP for all
$s=1,2,3...$. Notice that unless $s=1$, the dissipation parameter
$\sigma$ is not dimensionless. As we will see later, this redefinition
is convenient when boundaries are present. The goal is to define $Q_d$
through Eq. (\ref{dis2}) for $i=1\ldots N-1$, and to extend its
definition to $i=0,N$ such that the DP holds. Through a
straightforward expansion one gets
\begin{eqnarray*} 
&& (u,Q_du)_{\Sigma }^{[0,N]}
= \frac{\Delta x}{2} u_0 Q_du_0 + \sigma (\Delta x)^s
(u,D_+D_-u)^{[1,N-1]} + \frac{\Delta x}{2} u_N Q_du_N \\ &&= -\sigma
(\Delta x)^s \left( \| D_- u \|^{[1,N]} \right)^2 + \frac{\Delta
x}{2}u_0\left[Q_du_0 - 2\sigma (\Delta x)^{s-1}D_+u_0\right] \nonumber\\
&&+
\frac{\Delta x}{2}u_N\left[Q_du_N + 2\sigma (\Delta
x)^{s-1}D_-u_N\right] \; ,
\end{eqnarray*}
where the second equality comes from Eq. (\ref{prop2}) of appendix
\ref{App-BP}.

Certainly, there are many possible definitions of $Q_d$ at the
boundary points that imply the DP, Eq. (\ref{dissipation}). The
simplest one is
\begin{eqnarray}
&&Q_d u_0 = 2\sigma (\Delta x)^{s-1} D_+ u_0\, , \nonumber\\
&&Q_d u_j = \sigma (\Delta x)^s D_+D_- u_j, (j=1\ldots N-1), 
\label{eq:dissipation}\\
&&Q_d u_N = -2\sigma (\Delta x)^{s-1} D_- u_N\, .
\nonumber
\end{eqnarray}
Note that this dissipative term vanishes in the limit $\Delta x \to 0$
only if $s > 1$. On the other hand, in those cases, the amplification
factor of the high frequency modes depends on the resolution already
in the absence of boundaries. That is, consistency requires $s > 1$,
and in those cases the amplification factor converges to the
non-dissipative one as $\Delta x \to 0$. This means that the
dissipative operator just constructed cannot be expected to cure
difference schemes which are unstable. It can only be applied to
systems where the amplification factor does not become greater than
one in the limit of high resolution; that is, to schemes which are
already stable in the absence of dissipation. We now improve on this.

In the absence of boundaries, the standard way of correcting this is
by considering dissipative operators of the form (\ref{KO}) with $2r-1
\geq q$, which satisfy the DP and do not change the accuracy of a
scheme which uses $q$'th order accurate difference operators.
Therefore, we now look for corrections to the operator (\ref{KO}) at
and near the boundary points. For simplicity, let us consider the case
of a dissipative operator for an otherwise second order scheme; that
is, assume we are using difference operators of order two in the
interior, and first order at boundary points. Then, for $i=2\ldots
N-2$ we define $Q_d$ through equation (\ref{KO}) with $r=2$, that is,
we set
\begin{equation}
Q_d = -\sigma ( \Delta x)^s D_+^2 D_-^2 \; ,
\end{equation}
where we have redefined $\sigma$, as before.  The modification at and
near the boundary points is motivated by a calculation that is similar
to the one presented above:
\begin{eqnarray}
(u,Q_du)^{[0,N]} &= & \frac{1}{2}\Delta x u_0 Q_du_0 + 
\Delta x\; u_1 Q_d u_1 - 
\sigma (\Delta x)^s \left(u, D_+^2D_-^2u \right)_{[2,N-2]}  
\nonumber\\
&& +\; \Delta x u_{N-1}
Q_du_{N-1} + \frac{1}{2} \Delta x \; u_N Q_d u_N \\
&=& u_0\left(\frac{\Delta x}{2}Q_du_0 + \sigma (\Delta x)^{s-1}D_+^2u_0 \right)  
\nonumber\\
&& + \; u_1\left[\Delta xQ_du_1 + \sigma (\Delta x)^{s-1}
\left(D_+^2 \, \,  - 2D_+D_-\right)u_1\right] \nonumber\\
&& - \sigma (\Delta x)^s\left( \|D_+D_-u\|^{[1,N-1]} \right)^2 \nonumber\\
&& +\; u_{N-1}\left[ \Delta xQ_du_{N-1} + \sigma (\Delta x)^{s-1}
\left(D_-^2 \, \, -2D_+D_-\right)u_{N-1} \right]  \nonumber
\\&& +\; u_N\left(\frac{\Delta x}{2}Q_du_N + \sigma (\Delta
x)^{s-1}D_-^2 u_N \right)\; ,
\end{eqnarray}
where in the last equality we have used properties listed in appendix
\ref{App-BP}.

As before, there is more than one way of satisfying the DP, the
simplest one being: 
\begin{eqnarray}
Q_d u_0 &=& -2\sigma (\Delta x)^{s-2} D_+^2
u_0\nonumber \;, \\ Q_d u_1 &=& -\sigma (\Delta x)^{s-2} (D_+^2 -
2D_+ D_-)u_1\nonumber \;, \\
Q_d u_j &=& -\sigma (\Delta x)^s (D_+D_-)^2 u_j\; ,
\qquad j=2,\ldots,N-2 \nonumber \;, \\
Q_d u_{N-1} &=& -\sigma (\Delta x)^{s-2} (D_-^2 -2D_+D_-)u_{N-1}
\nonumber\;, \\
Q_d u_N &=& -2\sigma (\Delta x)^{s-2} D_-^2 u_N\;.
\label{eq:KOdiss21}
\end{eqnarray} 
For $s=3$ the high frequency modes are now damped in a resolution
independent way while the dissipative term vanishes in the limit
$\Delta x \to 0$. In this case, the dissipative operator constitutes a
third order correction in the interior and a first order one at the
boundary points and the two grid points next to it. Adding this
operator to the RHS of the equations does not affect the consistency
of the scheme, but reduces in one the order of accuracy at the grid
points $1$ and $N-1$ (recall that the difference scheme was already
first order at the boundary points, so the accuracy there is not
changed).

\subsection{Boundary conditions}

Finally, the maximally dissipative boundary conditions
(\ref{eq:MaxDiss}) are implemented by multiplying the RHS of
(\ref{eq:symhypsystem_discrete}) from left by an appropriate
projection operator \cite{olsson} $P$. For homogeneous boundary
conditions, $P$ is the projection on the space of grid functions that
satisfy the boundary conditions (\ref{eq:MaxDiss}) with $g=0$ which is
orthogonal with respect to the scalar product for which SBP holds.
The orthogonality of $P$ makes sure that the energy estimate is not
spoiled, and hence numerical stability still follows. Furthermore, if
$P$ is time-independent, a solution to the resulting semi-discrete
system automatically satisfies the boundary conditions. Inhomogeneous
boundary conditions are discussed in appendix \ref{App-BC}.

\section{SBP and dissipation for IBVP in domains with excised regions}
\label{Sect-excision}

In the previous section we restricted our discussions to computational
domains without inner boundaries. Now we extend those results to
non-trivial computational domains. To simplify the discussion we
restrict ourselves to the case where a single cubic box is cut out of
the computational domain. However, it is straightforward to see that
the difference and dissipative operators constructed in this section
can also be applied to a domain with multiple cubic boxes excised from
the computational domain, such that SBP and the dissipative property
(DP) are satisfied.

\subsection{Summation by parts}

We will only analyze the case in which the second order centered
difference operator $D_0$, defined in Eq. (\ref{Eq:SecondSD}), is used
in the interior. The possibility of constructing higher order accurate
operators is discussed in appendix \ref{App-FODiss}. The modification
of $D_0$ and its associated scalar product at the outer boundary
points has already been discussed in the previous section.
\begin{figure}[ht]
\begin{center}
\includegraphics*[height=8cm]{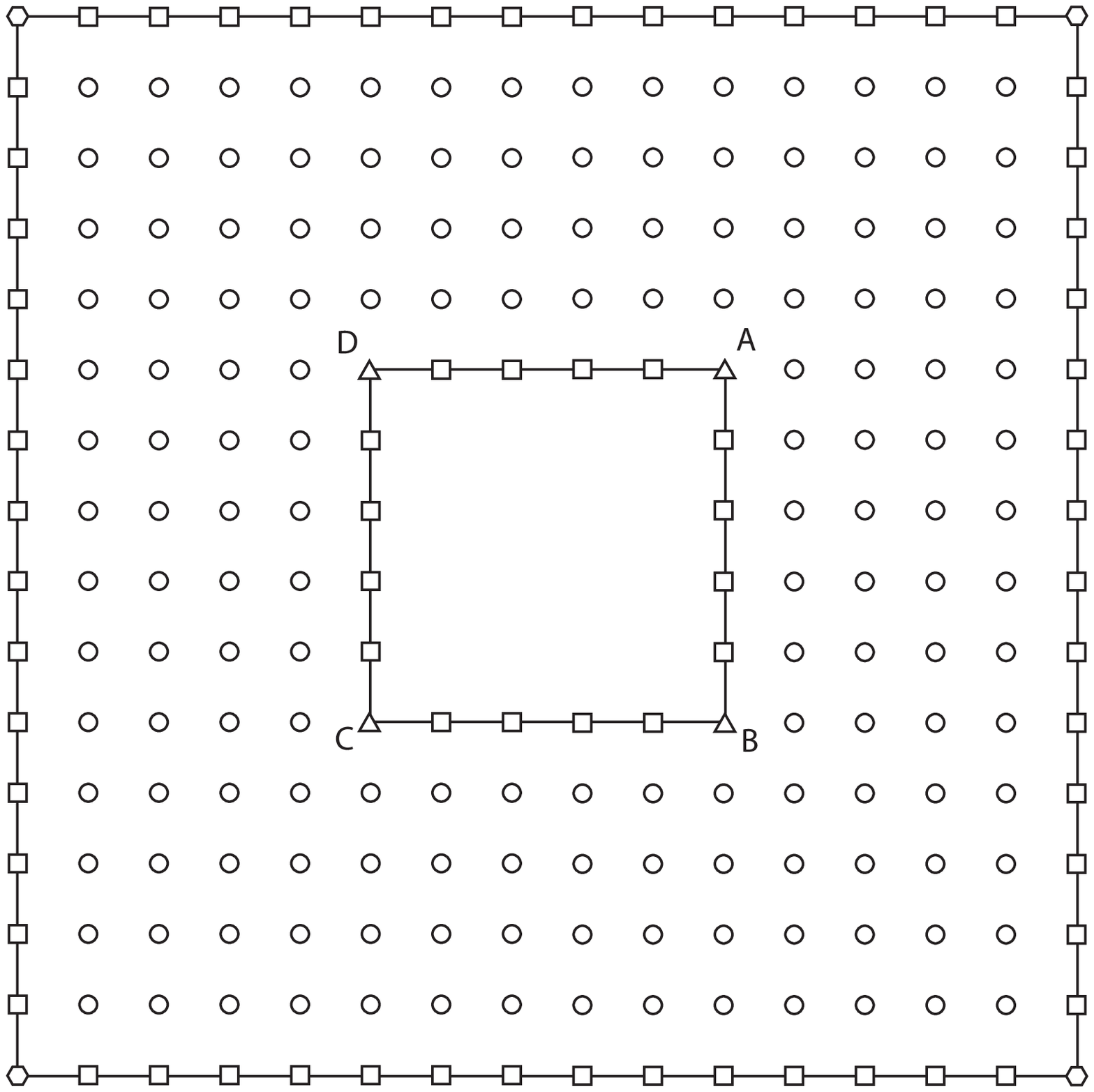}
\includegraphics*[height=8cm]{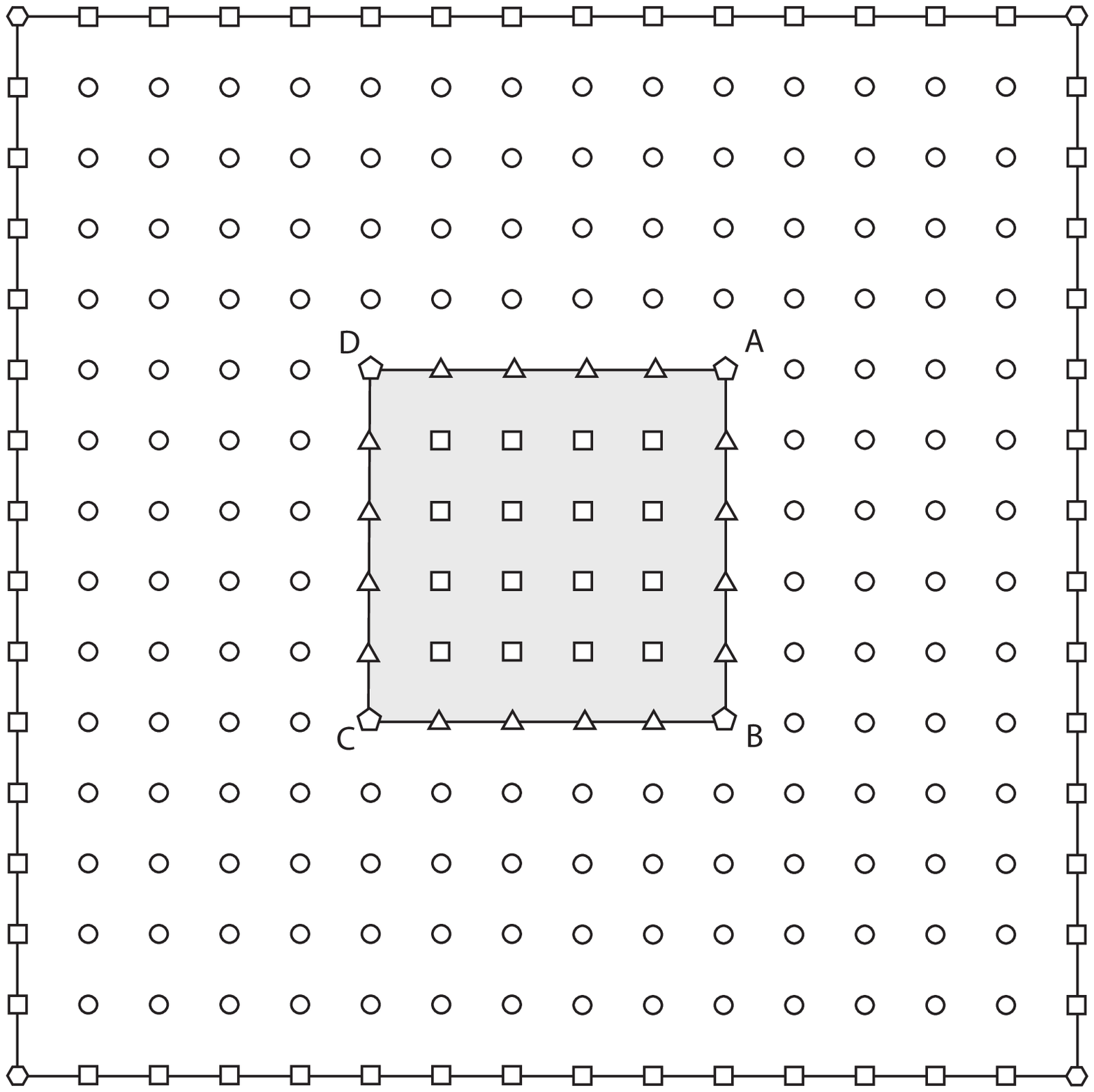}
\caption{The figure illustrates the set of grid points belonging to a
$z=\mbox{const.}$ plane that passes through the interior of the excised cube (left panel), 
or cuts the top or bottom of it (right panel, also shown in Fig.\ref{fig:leftright} ).  
The weights $\sigma_{ijk}$ that appear in the scalar product 
are 1 at interior grid points (circles), $1/2$ at boundary faces
(squares), $1/4$ at convex edges (hexagons), $1/8$ at convex vertices,
$3/4$ at concave edges (triangles) and $7/8$ at concave vertices (pentagons).
\label{fig:grid}} 
\end{center}
\end{figure}

In order to define the scalar product and the difference operators at
the excision boundary such that SBP holds, we restrict ourselves,  without loss of
generality, to an
arbitrary line which is parallel to the $x$ axis. For points on such a line we 
assume that the difference
operator in the $x$ direction depends only on neighboring points on
this line (as we will see next, this assumption suffices for our
purposes). In order to label the grid points on such a line we only
need the index $i$ and assume that there is an inner boundary at, say,
$i=0$. We will use centered differences at the neighbors $i=1$ and
$i=-1$ (this will also turn out to be sufficient) but leave the scalar
product at these points undefined for the moment. That is,
$\sigma_{ijk}$ at points $i=-1,0,1$ takes some values $\alpha$,
$\beta$, $\gamma$, respectively. Note that since the difference
operators at the outer boundaries have already been introduced in
section \ref{Sect-OB}, for the sake of clarity and without loss of
generality, we will now ignore this outer boundary and consider the
case where only an inner boundary is present. Therefore, in what
follows, the subindices range from $+ \infty$ to $- \infty$.

\begin{figure}[ht]
\begin{center}
\includegraphics*[height=2cm]{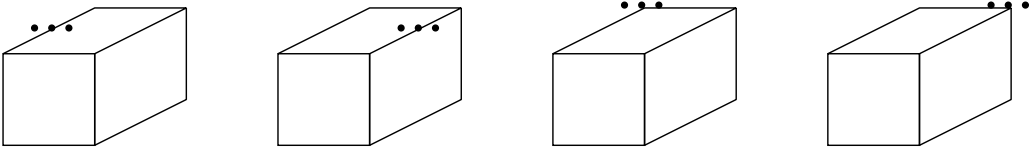}
\caption{Examples with interior points at the left and right of an edge, respectively, and to
the left and right of a vertex, respectively (sweeping from left to right in the figure).}
\label{fig:leftright}
\end{center}
\end{figure}

By splitting the scalar product in the different sub-domains one gets
(omitting the sum over $j$ and $k$)
\begin{eqnarray}
&&(u,Dv) + (v,Du) =
\alpha \left( (u,D_0v)^{(-\infty,-1]} + 
  (v,D_0u)^{(-\infty,-1]} \right) + \nonumber\\
&& \beta\Delta x \left( u_0 Dv_0 + v_0 Du_0 \right) + 
\gamma \left( (u,D_0v)^{[1,\infty)} +  (v,D_0u)^{[1,\infty)}  \right) \label{sbp} \\
        &=& \frac{\alpha }{2}(v_0 u_{-1} + u_0v_{-1}) 
+ \beta \Delta x \left(u_0 Dv_{0} + v_0 Du_0\right) 
- \frac{\gamma }{2} (u_0 v_1 + v_0u_1)\;. \nonumber
\end{eqnarray}
We now assume that the modified difference operator at $i=0$ has a
stencil of width three; that is, we assume that it has the form
\begin{equation}
Du_0 = \frac{q u_1 + r u_0 + s u_{-1}}{\Delta x} \;,
\label{gen_der}
\end{equation}
with $q$, $r$, $s$ to be determined later. Taylor-expanding
Eq. (\ref{gen_der}) in $\Delta x$ one obtains that in order for $Du_0$
to be consistent with $\partial_x$ the conditions
\begin{equation}
q+r+s=0 \; , 
\qquad
q-s=1
\label{consistency}
\end{equation}
must be satisfied. Inserting Eq. (\ref{gen_der}) and these conditions
in Eq. (\ref{sbp}) one sees that in order for SBP to hold, the mixed
terms of the form $v_0 u_{\pm 1}$ must vanish. This yields
\begin{equation}
\alpha + 2\beta s=0 \;\;\; , \;\;\; 2\beta q - \gamma=0 \; .\label{mixed}
\end{equation}
Equations (\ref{consistency}) and (\ref{mixed}) completely fix, at
$i=0$, the derivative operator (i.e.  the coefficients $q$, $r$ ,$s$),
and the scalar product (i.e., $\beta$):
\begin{eqnarray*}
\beta &=& \frac{\alpha+\gamma}{2} \label{beta}\;, \\
Du_0 &=& \frac{\gamma u_1 +(\alpha -\gamma)u_0 - \alpha
u_{-1}}{(\alpha + \gamma)\Delta x}\;. \label{nder}
\end{eqnarray*}

Notice that, unless $\alpha = \gamma$ (which in our case would
correspond to $i=0$ actually not being part of a boundary, as we will
now see), the modified difference operator (\ref{nder}) is only a
first order accurate approximation of $\partial_x$, as is the case
at the outer boundary points.

We now explicitly write down the scalar product and difference
operator just constructed for the different possible inner boundary
points:
\begin{itemize}
\item{\em Edges:}

By definition of an edge, one neighbor lies in the interior (and has
scalar product weight equal to $1$) and the other lies at a face (and
has scalar product weight equal to $1/2$). Therefore, Eq. (\ref{beta})
implies that $\beta = (1+1/2)/2 = 3/4$. Thus, the expressions for the
derivative operator (depending on the face's side) are,

\begin{table}[h]
\center
\begin{tabular}{||l|c|c||}\hline
interior points are (see Fig. \ref{fig:leftright}) & $(\alpha,\beta,\gamma)$ & $Du_0$ \\ \hline
to the left of face & $(1,3/4,1/2)$ & $\displaystyle\frac{ u_1 + u_0 - 2 u_{-1}}
{3 \Delta x}$  
\\ \hline 
to the right of face & $(1/2,3/4,1)$ & $\displaystyle
\frac{ 2u_1 - u_0 - u_{-1}}{3 \Delta x}$  
\\ \hline 
\end{tabular}
\end{table}

\item{\em Vertices:}

Similarly, at a vertex on neighbor lies in the interior (and has
scalar product weight equal to $1$), and the other lies at an edge
(and has scalar product weight equal to $3/4$), therefore $\beta =
(1+3/4)/2 = 7/8$ and the corresponding derivative operators are:

\begin{table}[h]
\center
\begin{tabular}{||l|c|c||}\hline
interior points are (see Fig. \ref{fig:leftright}) & $(\alpha,\beta,\gamma)$ & $Du_0$ \\ \hline
to the left of vertex & $(1,7/8,3/4)$ & 
$\displaystyle\frac{3u_1 + u_0 - 4u_{-1}}{7\Delta x}$  
\\ \hline 
to the right of vertex & $(3/4,7/8,1)$ & 
$\displaystyle\frac{4u_1 - u_0 - 3u_{-1}}{7\Delta x}$ 
 \\ \hline 
\end{tabular}
\end{table}

\end{itemize}

\subsection{Dissipative operator: second order form }

The dissipative operators computed in section \ref{artdiss} need to be
modified at the excision boundary. We again assume that the grid point
under consideration at the inner boundary is at index $0$ with $\sigma
_0=\beta$, and that at the neighbors, $\sigma_{-1} = \alpha$,
$\sigma_{1} = \gamma$. We expand the corresponding scalar product:
\begin{eqnarray}
(u,Q_du)_{\Sigma } &=& (u,Q_du)_{\Sigma }^{ (-\infty,-1]} +
(u,Q_du)_{\Sigma}^{(0,0)} + (u,Q_du)_{\Sigma }^{[1,\infty)} \nonumber\\
&=& \alpha (u,Q_du)^{(-\infty,-1]} + \beta (u,Q_du)^{(0,0)} + \gamma
(u,Q_du)^{[1,\infty)} \nonumber\\
&=& \alpha \sigma (\Delta x)^s (u,D_+D_-u)^{(-\infty,-1]} \nonumber\\
&& +\; \beta \Delta x\; u_0 Q_du_0  + \gamma
\sigma (\Delta x)^s (u,D_+D_-u)^{[1,\infty)} \nonumber\\
&=& -\sigma (\Delta x)^s\left[ \alpha \left(\|D_-u\|^{(-\infty,0]} \right)^2 
+ \gamma  \left( \|D_-u\|^{[1,\infty)} \right)^2 \right] \nonumber\\
&& +\; u_0\left[ \beta Q_d \Delta x + \sigma (\Delta x)^s\left(\alpha D_- - 
\gamma D_+ \right) \right] u_0 \; ,
\nonumber
\end{eqnarray}
where the last equality is due to Eq. (\ref{prop2}).

Therefore one (non-unique, but simple) possibility is to choose:
\begin{displaymath}
Q_du_0 = \frac{\sigma }{\beta} (\Delta x)^{s-1}( \gamma D_+ - \alpha D_-)u_0\;.
\end{displaymath}
in which case
\begin{displaymath}
(u,Q_du) =   -\sigma (\Delta x)^s\left[ \alpha \left( \|D_-u\|^{(-\infty,0]} \right)^2 + \gamma
  \left( \|D_-u\|^{[1,\infty)} \right)^2 \right] \;.
\end{displaymath}
Making the expressions for the dissipation explicit:
\begin{itemize}
\item{\em Edges:}

\begin{displaymath}
\begin{tabular}{||l|c|c||}\hline
interior points are (see Fig. \ref{fig:leftright}) & $(\alpha,\beta,\gamma)$ & $Q_du_0$ \\ \hline
to the left of face & $(1,3/4,1/2)$ & $ \frac{2}{3}\sigma (\Delta x)^{s-1}(D_+-2D_-)u_0$  \\ \hline 
to the right of face & $(1/2,3/4,1)$ & $ \frac{2}{3}\sigma (\Delta x)^{s-1}(2D_+-D_-)u_0$  \\ \hline 
\end{tabular}
\end{displaymath}

\item{\em Vertices:}

\begin{displaymath}
\begin{tabular}{||l|c|c||}\hline
interior points are (see Fig. \ref{fig:leftright}) & $(\alpha,\beta,\gamma)$ & $Q_du_0$ \\ \hline
to the left of face & $(1,7/8,3/4)$ & $ \frac{2}{7}\sigma (\Delta x)^{s-1}(3D_+-4D_-)u_0$  \\ \hline 
to the right of face & $(3/4,7/8,1)$ & $ \frac{2}{7}\sigma (\Delta x)^{s-1}(4D_+-3D_-)u_0$  \\ \hline 
\end{tabular}
\end{displaymath}

\end{itemize}

\subsection{Dissipative operator: fourth order form }

The modification in this case follows along similar lines:

\begin{eqnarray}
(u,Q_du)_{\Sigma } &=& (u,Q_du)_{\Sigma }^{(-\infty,-2]} +
 (u,Q_du)_{\Sigma }^{[-1,-1]} +
 + (u,Q_du)_{\Sigma}^{[0,0]}\nonumber\\
&& +\;  (u,Q_du)_{\Sigma}^{[1,1]} +
 (u,Q_du)_{\Sigma }^{[1,\infty)} \nonumber\\
&=& \alpha (u,Q_du)^{(-\infty,-2]}  +  \alpha (u,Q_du)^{[-1,-1]}  + 
\beta (u,Q_du)^{[0,0]} \nonumber\\
&+& \gamma (u,Q_du)^{[1,1]} + \gamma
 (u,Q_du)^{[2,\infty )} \nonumber\\
&=& -\alpha \sigma (\Delta x)^s \left( u,(D_+D_-)^2u \right)^{(\infty,-2]} +
\alpha \Delta xu_{-1}Q_du_{-1}\nonumber\\
&& +\; \beta \Delta x\; u_0 Q_du_0 + \gamma \Delta xu_1Q_du_1 - 
  \gamma \sigma (\Delta x)^s (u,D_+^2D_-^2u)^{[2,\infty)} \\
&=& -\sigma (\Delta x)^s\left[\alpha \left(\|D_+D_-u\|^{(-\infty,-1]} 
\right)^2 + \gamma
  \left( \|D_+D_-u\|^{-\infty,-1]} \right)^2\right ]\nonumber\\
&& +\; u_{-1}\left[\sigma (\Delta x)^{s-1}\alpha (2D_+D_--D_-^2)+\alpha \Delta
  x Q_d
\right]u_{-1}\nonumber\\
&& +\; u_0\left[ \beta \Delta xQ_d + \sigma (\Delta x)^{s-1}(\alpha D_-^2+\gamma D_+^2)
\right]u_0\nonumber\\
&& +\; u_1\left[\gamma \Delta x Q_du_1 + \sigma (\Delta x)^{s-1}\gamma
  (D_+^2-2D_+D_-) \right]\; ,\nonumber
\end{eqnarray}
where the last equality is due to Eq. (\ref{prop3}).

Therefore one possible choice is
\begin{eqnarray*}
Q_du_{-1} &=& -\sigma (\Delta x)^{s-2} (D_-^2 - 2D_+D_-)u_{-1}\;, \\
Q_du_0 &=& -\frac{\sigma (\Delta x)^{s-2}}{\beta} (\alpha D_-^2 + \gamma
D_+^2)u_0\;, \\
Q_du_{1} &=& -\sigma (\Delta x)^{s-2} (D_+^2 - 2D_+D_-)u_{1} \;.
\end{eqnarray*}
The case $s=3$ should be the preferred one, since then the
amplification factor does not depend on resolution.

Making the expressions for the dissipation explicit:

\begin{itemize}
\item{\em Edges:}

\begin{displaymath}
\begin{tabular}{||l|c|c||}\hline
interior points are (see Fig. \ref{fig:leftright}) & $(\alpha,\beta,\gamma)$ & $Q_d u_0$ \\ \hline
to the left of face & $(1,3/4,1/2)$ & $ -\frac{2}{3}\sigma (\Delta x)^{s-1}(2D_-^2+D_+^2)u_0$  \\ \hline 
to the right of face & $(1/2,3/4,1)$ & $ -\frac{2}{3}\sigma (\Delta x)^{s-1}(D_-^2+2D_+^2)u_0$  \\ \hline 
\end{tabular} 
\end{displaymath}

\item{\em Vertices:}

\begin{displaymath}
\begin{tabular}{||l|c|c||}\hline
interior points are (see Fig. \ref{fig:leftright}) & $(\alpha,\beta,\gamma)$ & $Q_du_0$ \\ \hline
to the left of vertex & $(1,7/8,3/4)$ & $ -\frac{2}{7}\sigma (\Delta x)^{s-1}(4D_-^2+3D_+^2)u_0$  \\ \hline 
to the right of vertex & $(3/4,7/8,1)$ & $ -\frac{2}{7}\sigma (\Delta x)^{s-1}(3D_-^2+4D_+^2)u_0$  \\ \hline 
\end{tabular}
\end{displaymath}

\end{itemize}

\section{Conclusions}
\label{Sect-Conclusions}

We have presented a set of 3D operators satisfying summation by parts
which can be used in non-trivial domains. The use of these operators
to numerically implement first order symmetric hyperbolic systems
guarantee stability of the semi-discrete system. Additionally, the use
of appropriate time integrators and a consistent treatment of the
boundary provide a systematic way to achieve a stable implementation.
Furthermore, in order to rule out artificial growth, we have defined
dissipative operators which do not spoil the energy growth estimates.

The usefulness of these techniques will be highlighted in their
application to different problems elsewhere. In particular, these
applications include the evolution of scalar and electromagnetic
fields propagating in black hole backgrounds in $3D$ \cite{lsulong2};
scalar fields in $2D$ with a moving black hole \cite{davegioel};
bubble spacetimes which require evolving the 5D Einstein equations in
the presence of symmetries \cite{bubble1}, the construction of schemes
capable to deal with the axis of symmetry in $2D$, axisymmetric
scenarios \cite{olivieraxis} and 3D numerical simulations of single
black holes in full nonlinear GR \cite{hyperGR}.

\section{Acknowledgments}

We would like to thank H.O. Kreiss, D. Neilsen, P. Olsson, J. Pullin,
and E. Tadmor for very useful discussions, comments and suggestions
throughout this work. This work was supported in part by grants
NSF-PHY-9800973, NSF-PHY-0244335, NSF-INT-0204937, PHY99-07949,
NASA-NAG5-13430, by funds of the Horace Hearne Jr. Institute for
Theoretical Physics, and by Agencia C\'ordoba Ciencia, CONICET, and
SeCyT UNC. LL has been partially supported by the Sloan Foundation.

\appendix

\section{Basic properties of finite difference operators}
\label{App-BP}

In this appendix the definition and some properties of the first order
\begin{displaymath}
D_+ u_j = \frac{u_{j+1}-u_j}{\Delta x} \;\;\; , \;\;\;
D_- u_j = \frac{u_j - u_{j-1}}{\Delta x} \;, 
\end{displaymath}
and second order accurate
\begin{displaymath}
D_0 u_j = \frac{u_{j+1} - u_{j-1}}{2\Delta x} 
\end{displaymath}
finite difference operators are given. The proofs of the following
statements can be found in Ref. \cite{GKO}, for instance. One can show
that with respect to the scalar product and norm
\begin{displaymath}
(u,v)^{[r,s]} \equiv \sum_{j=r}^s u_jv_j \Delta x, 
\qquad \left(\|u\|^{[r,s]} \right)^2 = (u,u)^{[r,s]}\;, 
\end{displaymath}
the following properties hold:
\begin{eqnarray*}
(u,D_+v)^{[r,s]} &=& - (D_-u,v)^{[r+1,s+1]} + u_j v_j |_r^{s+1}\\\
&=& -(D_+u,v)^{[r,s]} - h(D_+u,D_+v)_{[r,s]} + u_jv_j|_r^{s+1}\;, \\
(u,D_-v)^{[r,s]} &=& - (D_+u,v)_{[r-1,s-1]} + u_j v_j |_{r-1}^s \\
&=& -(D_-u,v)_{[r,s]} + h(D_-u,D_-v)_{[r,s]} + u_jv_j|_{r-1}^s\;, \\
(u,D_0v)^{[r,s]} &=& -(D_0u,v)_{[r,s]} + \frac{1}{2}(u_jv_{j+1} +
u_{j+1}v_j)|_{r-1}^s\;.
\end{eqnarray*}
The following equalities are needed in the derivation of the modified
derivative and dissipative operators near the boundary
\begin{eqnarray}
(u,D_0u)^{[r,s]} &=& \frac{1}{2}(u_ju_{j+1})|_{r-1}^s\;,
\\
(u,D_+D_-u)^{[r,s]} &=& -\left( \|D_-u\|^{[r,s+1]} \right)^2 + u_{s+1}D_-u_{s+1} - 
u_{r-1}D_+ u_{r-1} \; ,
\label{prop1}\\
(u,(D_+D_-)^2u)^{[r,s]} 
&=& \left( \|D_+D_-u\|^{[r-1,s+1]} \right)^2 + \frac{u_{s+1}}{\Delta x}\left(2D_+D_- -
D_-^2\right)u_{s+1} 
\nonumber\\
&& - \frac{u_{s+2}}{\Delta x}D_-^2 u_{s+2}+\frac{u_{r-1}}{\Delta x}\left(2D_+D_- - 
D_+^2\right)u_{r-1} - \frac{u_{r-2}}{\Delta x}D_+^2 u_{r-2} \; .
\label{prop3} 
\end{eqnarray}
A sketch of the calculation that leads to the last equality is:
\begin{eqnarray}
(u,(D_+D_-)^2u)^{[r,s]} &=& \left( \|D_+D_-u\|^{[r,s]} \right)^2 \nonumber\\
&& + \left. \left( u_j D_-D_+D_-
u_j- D_-u_j D_+D_-u_j\right)\right|_r^{s+1}
\label{prop2}\\
&=& ( \|D_+D_-u\|^{[r,s]} )^2 +
\frac{1}{h}\left. (u_{j-1}D_+^2u_{j-1} - u_jD_-^2u_j)\right|_r^{s+1}
\nonumber\\
&=& \left( \|D_+D_-u\|^{[r-1,s+1]} \right)^2 + \frac{u_{s+1}}{\Delta x}\left(2D_+D_- -
D_-^2\right)u_{s+1}
\nonumber\\
&&- \frac{u_{s+2}}{\Delta x}D_-^2 u_{s+2}+\frac{u_{r-1}}{\Delta x}\left(2D_+D_- - 
D_+^2\right)u_{r-1} - \frac{u_{r-2}}{\Delta x}D_+^2 u_{r-2}\;.
\nonumber
\end{eqnarray}

\section{On the existence of certain class of fourth order difference operators 
satisfying SBP when excised regions are present}
\label{App-FODiss}

Here we analyze the question of whether one can construct high order
difference operators satisfying SBP in a domain with excised cubic
regions. More precisely, we seek difference operators that are fourth
order in the interior and, as in the rest of the article, we assume that
the 3D operator is inherited by a 1D one with diagonal scalar product,
that the operator approximating, say, $\partial_x$, depends only on
points to the left or the right. Unfortunately the calculation below
shows that such an operator cannot be constructed. However, a difference
operator that satisfies SBP and is fourth order accurate almost everywhere
in the interior could be obtained by means of a different approach
where one decomposes the domain into cubes \cite{olsson_private}.

\subsection{Fourth order accurate operators for domains without excised regions}
 
The finite difference operator
\begin{equation}
(Q u)_j = D_0\left(I-\frac{(\Delta x)^2}{6}D_+D_-\right) u_j =
\frac{-u_{j+2}+8u_{j+1}-8u_{j-1}+u_{j-2}}{12\Delta x}
\label{D4}
\end{equation}
is a fourth order approximation of $d/dx$. Strand \cite{strand} showed
that there exists a unique second order accurate modification of this
operator near the boundaries, given by
\begin{eqnarray}
(Qu)_0 &=& (-48u_0+59u_1-8u_2-3u_3)/(34\Delta x)\;, \nonumber\\
(Qu)_1 &=& (u_2-u_0)/(2\Delta x)\nonumber \;, \\
(Qu)_2 &=& (8u_0-59u_1+59u_3-8u_4)/(86\Delta x)\nonumber\;, \\
(Qu)_3 &=& (3u_0-59u_2+64u_4-8u_5)/(98\Delta x)\nonumber \;, \\
(Qu)_{N-3} &=& (8u_{N-5}-64u_{N-4}+59u_{N-2}-3u_N)/(98\Delta
x)\nonumber \;, \\
(Qu)_{N-2} &=& (8u_{N-4}-59u_{N-3}+59u_{N-1}-8u_N)/(86\Delta
x)\nonumber \;, \\
(Qu)_{N-1} &=& (u_{N}-u_{N-2})/(2\Delta x)\nonumber \;,\\
(Qu)_N &=& (3u_{N-3}+8u_{N-2}-59u_{N-1}+48u_N)/(34\Delta x)\label{eq:Strand42}
\end{eqnarray}
that satisfies SBP with respect to a diagonal scalar product, 
with 
\begin{displaymath}
\sigma_i = \left\{\frac{17}{48},\frac{59}{48},\frac{43}{48},
\frac{49}{48}, 1, 1, \ldots, 1, 1, \frac{49}{48}, \frac{43}{48},
\frac{59}{48}, \frac{17}{48} \right\}.
\end{displaymath}

Next we discuss the question of whether one can modify this operator near inner 
boundaries, for the case of excised cubic regions, in a way such that SBP holds.

\subsection{Modification of the fourth order operator near an inner boundary}
Let us assume that a 2D domain has a concave corner at
the grid point $(0,0)$, as is the case of an edge at the inner boundary of our domain.  The scalar product near this point will have
the following structure
\begin{equation}
\begin{array}{cccccccccc}
& \vdots & \vdots & \vdots & \vdots & \vdots & \vdots  &\vdots & \vdots & \\
\ldots & 1 & 1 & 1 & 1 & 1 & 1 & 1 & 1 & \ldots\\
\ldots & 1 & 1 & 1 & 1 & 1 & 1 & 1 & 1 & \ldots\\
\ldots & \frac{49}{48} & \frac{49}{48} & \sigma_{03} & \sigma_{13} & \sigma_{23} & \sigma_{33} & 1 &  1 &\ldots\\
\ldots & \frac{43}{48} & \frac{43}{48} & \sigma_{02} & \sigma_{12} & \sigma_{22} & \sigma_{32} & 1 &  1 &\ldots\\
\ldots & \frac{59}{48} & \frac{59}{48} & \sigma_{01} & \sigma_{11} & \sigma_{21} & \sigma_{31} & 1 & 1 & \ldots\\
\ldots & \frac{17}{48} & \frac{17}{48} & \sigma_{00} & \sigma_{10} & \sigma_{20} & \sigma_{30} & 1 & 1 & \ldots\\
\ldots & 0             & 0             & \frac{17}{48} & \frac{59}{48} & \frac{43}{48} & \frac{49}{48} & 1 & 1 & \ldots\\
\ldots & 0             & 0             & \frac{17}{48} & \frac{59}{48} & \frac{43}{48} & \frac{49}{48} & 1 & 1 & \ldots\\
& \vdots & \vdots & \vdots & \vdots & \vdots & \vdots  &\vdots & \vdots & 
\end{array} \label{scheme}
\end{equation}
The weights $\sigma_{ij}$ for $0 \le i,j \le 3$ are unknown.  We know
that they must be positive and symmetric $\sigma_{ij} = \sigma_{ji}$.
We also need to compute a second order accurate difference operator.

Consider one of the rows with $1 \le j \le 3$. Their contribution to
the scalar product $(u,Du)_h$ must be zero, as there are no boundary
terms at the continuum. We have
\begin{equation}
(u,Qu) = \sigma_j (u,D^{(4)}u)^{(-\infty,-1]} + \Delta x \sum_{i=0}^3 \sigma_{ij}
u_i (Du)_i + (u,D^{(4)}u)^{[4,+\infty) } \;, \label{rowj}
\end{equation} 
where $D^{(4)}$ is the fourth order accurate difference operator
defined in Eq.~(\ref{D4}).

The contribution from the first and last term of Eq.~(\ref{rowj}) can
be calculated using the properties listed in appendix \ref{App-BP}.
In general, one has that
\begin{equation}
(u,D^4 u)^{[r,s]} = \frac{1}{12} \left.\left( 8 u_{j+1}
u_j - u_{j+2} u_j - u_{j+1}u_{j-1}\right)\right|_{r-1}^s\;.
\end{equation}
We want to solve 
\begin{eqnarray}
&&\frac{\sigma_j}{12}(8u_0u_{-1} - u_1 u_{-1} -u_0u_{-2}) +  
\Delta x \sum_{i=0}^3 \sigma_{ij}
u_i (Du)_i \nonumber\\
&&- \frac{1}{12} (8u_4u_3 - u_5u_3 -u_4u_2) = 0
\label{rowjbexp}
\end{eqnarray}
for the coefficients of the modified difference operator $(Du)_i$ and the
weights of the scalar product $\sigma_{ij}$ for $i=0,1,2,3$.

The operator $(Du)_i$ can be at most a $7$ point stencil, i.e.
\begin{displaymath}
(Du)_i = \frac{\gamma_3 u_{i+3} +\gamma_2 u_{i+2} + \gamma_1 u_{i+1} 
+ \gamma_0 u_{i} + \gamma_{-1} u_{i-1} + \gamma_{-2} u_{i-2} 
+ \gamma_{-3} u_{i-3} }{\Delta x} \;.
\end{displaymath}
A larger stencil would give rise to terms in (\ref{rowj}) that would
not cancel.

By inspection of  Eq. (\ref{rowjbexp}) carefully, one sees that there are many
coefficients of the modified difference operators that must vanish.
This reduces the number of parameters from $7\times4=28$ to $6$.
Eq.~(\ref{rowjbexp}) becomes
\begin{eqnarray*}
0 &=& \frac{\sigma_j}{12}(8u_0u_{-1} - u_1 u_{-1} -u_0u_{-2}) \\
&& + \sigma_{0j} u_0 \left[ -\frac{3}{8} (a_0 + b_0)u_3 + a_0 u_2 +
\left(\frac{1}{2} - \frac{3}{4} a_0 + \frac{5}{4} b_0\right)u_1 \right.\nonumber\\
&&\left.+
\left( - \frac{1}{2} + \frac{1}{8} a_0 - \frac{15}{8}b_0\right) u_{-1}
+ b_0 u_{-2} \right]\\
&& + \sigma_{1j} u_1 \left[ a_1 u_3 + \left( \frac{1}{2} - 2a_1\right)
u_2 + \left(-\frac{1}{2}+2a_1\right) u_0 - a_1 u_{-1} \right]\\
&& + \sigma_{2j} u_2 \left[ a_2 u_4 + \left( \frac{1}{2} - 2a_2\right)
u_3 + \left(-\frac{1}{2}+2a_2\right) u_1 - a_2 u_{0} \right]\\
&& + \sigma_{3j} u_3 \left[ a_3 u_5 + \left(\frac{1}{2} -\frac{15}{8}
a_3 +\frac{1}{8} b_3\right) u_4 + \left( -\frac{1}{2} + \frac{5}{4}
a_3 - \frac{3}{4} b_3\right) \right.\nonumber\\
&&\left.+ b_3 u_1 -\frac{3}{8}(a_3 + b_3)u_0 \right]\\
&&- \frac{1}{12} (8u_4u_3 - u_5u_3 -u_4u_2) \;. 
\end{eqnarray*}

Unless $\sigma_j = 1$ there is no solution; to see this it is
sufficient to look at the coefficients of $u_0u_{-2}$, $u_1u_{-1}$, $u_2u_0$,
$u_3u_0$, $u_3u_1$, $u_4u_2$ and $u_5u_3$. 

On the other hand, $\sigma _j =1$ is in contradiction with the structure of the scalar product under consideration, 
see (\ref{scheme}).

\section{Courant Limits}
\label{App-CFL}
\subsection{Courant limits}

When discretizing a hyperbolic system of partial differential
equations with an explicit scheme the Courant-Friedrich-Lewy (CFL)
condition has to be satisfied in order to have numerical stability.
Below we obtain necessary and sufficient conditions for the numerical
stability of a 3D wave equation, using a standard von Neumann
analysis. In particular, we want to determine what is the maximum
value of $\lambda = \Delta t/\Delta x$ that one can use if artificial
dissipation is added to the RHS.  One can use this information as
guide for choosing the Courant factor in more general situations.

We start by considering the scalar advective equation.

\subsubsection{Scalar advective equation}

Consider the advective equation $u_t = a u_x$, with $a$ a real
constant. If the spatial derivative is discretized using a second
order centered differencing operator and the resulting semi-discrete
system is integrated using a Runge--Kutta time integrator of order
$p$, the solution at a time step $n+1$ can be expressed in terms of
the solution at the previous time step $n$ as
\begin{equation}
u^{n+1}_k = \sum_{s=0}^p \frac{1}{s!}(a \Delta t D_0 )^s u^n_k
\label{eq:scheme_advective} \;. 
\end{equation}
This difference scheme (\ref{eq:scheme_advective}) is stable if and
only if it satisfies the von Neumann condition.  The Courant limit is
$a\lambda \le \sqrt{3}$ for the $p=3$ case (third order Runge-Kutta)
and $a\lambda \le 2\sqrt{2}$ for the $p=4$ case (fourth order
Runge-Kutta).

\begin{figure}[ht]
\begin{center}
\includegraphics*[height=8cm]{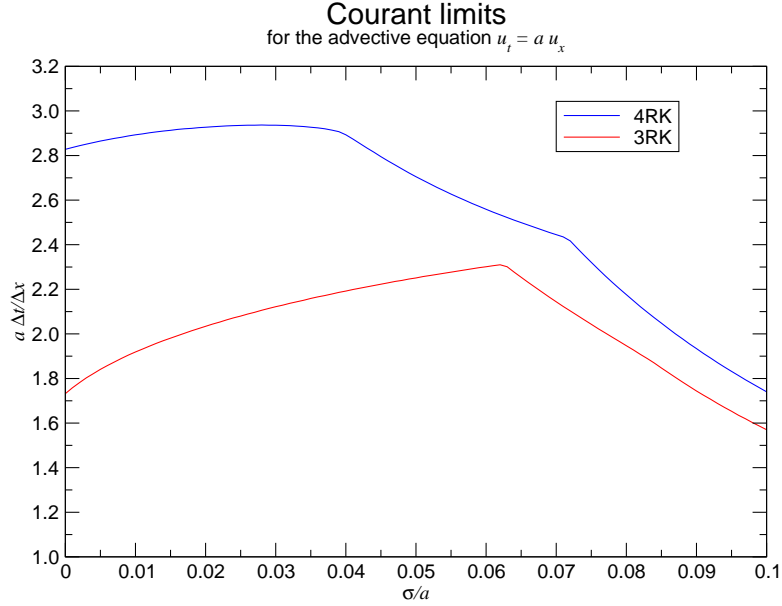}
\caption{The maximum value of the Courant factor that satisfies the von
Neumann condition is plotted as a function of the dissipation
parameter $\sigma$. \label{fig:courantlimits_1d}} 
\end{center}
\end{figure}

These limits change when artificial dissipation is added to the RHS.
The fully discrete system becomes
\begin{equation}
u^{n+1}_k = \sum_{s=0}^p \frac{1}{s!}\left(a \Delta t D_0 - \sigma \Delta t
(\Delta x)^3 (D_-D_+)^2\right)^s u^n_k \label{eq:scheme_advective_dissip}\;.
\end{equation}
The amplification factor depends on the parameters
$\tilde{\lambda} = \lambda a$ and $\tilde{\sigma} = \sigma/a$.  
If, for a given value of $\tilde{\sigma}$, we numerically compute the value of
$\tilde{\lambda}$ beyond which the amplification factor becomes in
magnitude greater than one for some frequency, we obtain the plot
in figure \ref{fig:courantlimits_1d}.

\subsubsection{3D wave equation}

\begin{figure}[ht]
\begin{center}
\includegraphics*[height=8cm]{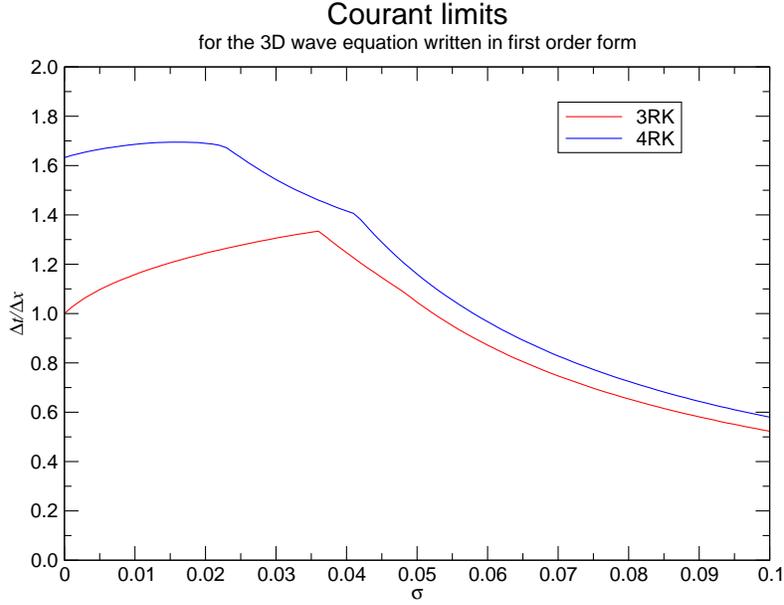}
\caption{The maximum value of the Courant factor for the stability of
the difference scheme approximating the first order 3D wave equation as a
function of the dissipation parameter $\sigma$. \label{fig:courantlimits_3d}} 
\end{center}
\end{figure}

Consider now the 3D wave equation $\phi_{tt} =
\phi_{xx}+\phi_{yy}+\phi_{zz}$ written in first order form. As before, we use
the second order centered difference operator to approximate the
spatial derivatives and $p$-th order Runge-Kutta.  The finite
difference scheme is 
\begin{displaymath}
\bm{u}^{n+1}_{ijk} = \sum_{s=0}^p \frac{1}{s!}  ( \Delta t Q )^s \bm{u}^n_{ijk}
\label{eq:3Dscheme}
\end{displaymath}
where $\bm{u} = (\phi_t,\phi_x,\phi_y,\phi_z)^T$, 
\begin{eqnarray*}
Q &=& A^{(x)}
D^{(x)}_0 + A^{(y)} D^{(y)}_0 + A^{(z)} D^{(z)}_0 
-\sigma  (\Delta x)^3 (D^{(x)}_+D^{(x)}_-)^2 \nonumber\\
&&-\sigma  (\Delta y)^3 (D^{(y)}_+D^{(y)}_-)^2 
-\sigma  (\Delta z)^3 (D^{(z)}_+D^{(z)}_-)^2 \;.
\end{eqnarray*}
We assume that $\Delta x= \Delta y = \Delta z$ and set $\lambda =
\Delta t / \Delta x$. In Fourier space 
the difference scheme becomes
\begin{displaymath}
\hat{\bm{u}}^{n+1}(\xi,\eta,\zeta) = G(\xi,\eta,\zeta)
\hat{\bm{u}}^{n}(\xi,\eta,\zeta)
\end{displaymath}
and the amplification matrix is given by
\begin{displaymath}
G(\xi,\eta,\zeta) = \sum_{s=0}^p \frac{1}{s!}
\left( \begin{array}{cccc}
S    & 0 & 0 & i R_x \\
0    & S & 0 & i R_y \\
0    & 0 & S & i R_z \\
iR_x & iR_y & iR_z & S
\end{array} \right)^s
\end{displaymath}
where $R_x = \lambda \sin(\xi)$, $R_y = \lambda \sin(\eta)$, $R_z =
\lambda \sin(\zeta)$ and $S = -16\sigma\lambda(\sin(\xi/2)^4 +
\sin(\eta/2)^4 + \sin(\zeta/2)^4)$.  The maximum values for the Courant factor are
plotted in figure \ref{fig:courantlimits_3d}.

\section{Boundary conditions}
\label{App-BC}

Here we discuss how to discretize inhomogeneous boundary conditions of
the type (\ref{eq:MaxDiss}). Let $P$ be the orthogonal projector onto
the space of grid functions that satisfy the homogeneous version of
the boundary conditions (\ref{eq:MaxDiss}). By  {\it orthogonal} we mean
that $P$ is Hermitian with respect to the scalar product for which SBP
holds. Eq. (\ref{eq:symhypsystem}) is discretized in the following
way:
\begin{equation}
\partial_t v = P A D v + B v + (I - P)\partial_t g\, ,
\label{eq:semidiscrete}
\end{equation}
where $I$ denotes the identity matrix on the space of grid functions.
Numerical stability for the resulting semi-discrete scheme can be
shown under additional assumptions on the boundary data (see
\cite{olsson}).

The construction of the RHS in (\ref{eq:semidiscrete}) proceeds as
follows:
\begin{itemize}
\item Discretize the partial derivatives using a difference operators
  that satisfies SBP.
\item At each $x_{ijk} \in \Gamma$ 
\begin{itemize}
\item Calculate the outward pointing unit normal $n$ and let 
$A_n= A^in_i$ denote the boundary matrix. Note that the normal at
edges and vertices needs special consideration.  
In the next subsection it is discussed how this normal is to be defined
in these cases.
\item Multiply the RHS by $Q^T$, where $Q$ is the (orthogonal) matrix
that diagonalizes $A_n$, i.e.~$Q^T A_nQ = {\rm
diag}(\Lambda_+,\Lambda_-,0)$, where $\Lambda_{\pm}$ is a diagonal
positive (negative) definite matrix.  Let $(W_+, W_-,W_0)^T = Q^T
ADv$;
\item Apply $P$. Since it is Hermitian with respect to the scalar
product for which SBP holds, this projector is non-trivial in the
sense that, in general, it overwrites the ingoing {\em and outgoing}
modes. Consider, for simplicity, the case in which there is one
ingoing and one outgoing mode (or more generally, the case in which
the coupling matrix between ingoing and outgoing modes is diagonal).
Then, we define
\begin{eqnarray*}
W_+^{(new)} &=& \frac{S}{1+S^2}(S W_+^{(old)} + W_-^{(old)}) + \frac{1}{1+S^2}\dot{g}\\
W_-^{(new)} &=& \frac{1}{1+S^2}(S W_+^{(old)} + W_-^{(old)}) - \frac{S}{1+S^2}\dot{g}
\end{eqnarray*}
where $S$ is the coupling that appears in the boundary condition $w_+
= S w_- + g$. Note that only when $S=0$ does the outgoing mode remain
unchanged, otherwise it is overwritten. In the continuum limit, this
overwriting is, of course, consistent with the partial differential
equation.
\end{itemize}
\item Go back to the primitive variables $v$ by multiplying by $Q$.
\end{itemize}

\subsection{Corners, edges and vertices}

It is useful to have stability results that allow for cubic domains,
since they are used quite often. The corners in 2D or edges and
vertices in 3D deserve special attention. In particular, the normal to
the boundary is not defined there. The stability of the whole scheme
is affected by the treatment of these points. In Ref. \cite{olsson} it
is discussed how to control (i.e. how to achieve numerical
stability) the boundary terms that appear after SBP at non-smooth
parts of the boundary. Here we will summarize some aspects of this
treatment.

For simplicity, let us assume that the domain is a 2D square $\Omega =
[0,1]\times [0,1]$ and that the partial differential equation is a
symmetric hyperbolic, constant-coefficients one with no principal
terms,
\begin{equation}
\partial_t u = A^x \partial_x u + A^y \partial_y u   \label{sym2d}
\end{equation}
where $A^x$ and $A^y$ are constant symmetric matrices. If we define
the energy norm to be ${\mathcal E} = \int_{\Omega} u^T u \,dx\,dy$,
the time derivative of the energy norm will be given the boundary
terms
\begin{equation}
\frac{d}{dt}{\mathcal E} = \int_0^1 u^T A^x u|_{x=0}^{x=1} \,dy + \int_0^1 u^T
A^yu|_{y=0}^{y=1} \,dx\;.
\label{eq:contestimate}
\label{edotcont}
\end{equation}
In particular, in these boundary terms, there is no contribution due to
corners, since they constitute a set of measure zero.

The semi-discrete energy estimate obtained by discretizing the RHS
with a difference operator satisfying SBP is simply the discrete
version of (\ref{eq:contestimate}):
\begin{eqnarray}
\frac{d}{dt}E &=& \sum_{j=0}^{N_y}\Delta y \sigma_j 
( u^T_{N_xj}A^xu_{N_xj} - u^T_{0j} A^x
u_{0j} ) \nonumber\\
&& + \sum_{i=0}^{N_x} \Delta x \sigma_i 
( u^T_{iN_y}A^yu_{iN_y} - u^T_{i0} A^y u_{i0} )\;.
\label{eq:semidiscrestimate}
\end{eqnarray}
One needs to prescribe boundary conditions to the semi-discrete system
at the corners. There, the unit vector $n$ is not defined. In
\cite{olsson} it is shown how to define $n$ so that a bound to the
semi-discrete energy estimate can be given. Essentially, the idea is
that by looking at the contribution coming from the corner $(x,y) =
(0,0)$ in (\ref{eq:semidiscrestimate}), which is given by,
\begin{displaymath}
-\sigma_0 u^T_{00} \left( \Delta y A^x + \Delta x A^y\right) u_{00}
\end{displaymath}
(note that the ``cross-terms'' $\Delta y A^x$ and $\Delta x A^y$ are
{\em not} a typo but do result from a non-trivial contribution to the
discrete energy), we see that we need to control the positive speed
characteristic variables in the effective direction $n = - (\Delta
y,\Delta x)/\Delta$, where $\Delta = ((\Delta x)^2+(\Delta y)^2)^{1/2}$.
For uniform grids with $\Delta x = \Delta y$ this effective unit
vector at the corners of $\Omega$ is
\begin{eqnarray}
&&m(0,0) = (-1,-1)/\sqrt{2}\, , \;\;\;
  m(1,0) = (+1,-1)/\sqrt{2}\, ,\\
&&m(0,1) = (-1,+1)/\sqrt{2}\, , \;\;\; 
  m(1,1) = (+1,+1)/\sqrt{2}\, . \nonumber
\end{eqnarray}
This is equivalent to providing boundary conditions as if the normal
was at 45 degrees with respect to the faces of the cube. Which data to
give might be completely or partially determined by compatibility
conditions. As discussed before, the way in which the boundary
conditions are imposed in \cite{olsson}, including at corners, is
through a non-trivial orthogonal projector.

\subsection{A brief discussion about other boundary conditions at corners}

One might wonder whether or not giving boundary conditions at corners
along the normal to one of the two faces that define the corner yields
an energy estimate. That is, if controlling characteristic fields in
one of the directions (say, the term $u^T \Delta y A^x u$ ) implies
that the other one is also under control because of some compatibility
conditions. The general answer is {\em no}, but it might work in some
particular cases. Let us illustrate this with one example.

Consider the 2D symmetric hyperbolic system
\begin{displaymath}
\partial _t \left(
\begin{array}{c}
u \\ v
\end{array}
\right)
=
\left(
\begin{array}{cc}
1 & 0 \\ 0 & -1
\end{array}
\right) \partial_x \left(
\begin{array}{c}
u \\ v
\end{array}
\right) + 
\left(
\begin{array}{cc}
0 & 1 \\ 1 & 0
\end{array}
\right) \partial_y \left(
\begin{array}{c}
u \\ v
\end{array}
\right)
\end{displaymath}
in the domain $x > 0$, $y > 0$. The characteristic fields in the
$x$ direction are $u$ (eigenvalue $1$) and $v$ (eigenvalue $-1$). In
the $y$ direction they are $u+v$ (eigenvalue $1$) and $u-v$
(eigenvalue $-1$). As boundary conditions in the $x=0$ and $y=0$
boundaries use $v=su$ and $u-v = r (u+v)$, respectively, with
$|r|,|s|\leq 1$.

Defining the energy 
\begin{displaymath}
{\mathcal E}= \frac{1}{2}\int_0^{\infty} \int_0^\infty ( u^2+v^2 ) dx dy,
\end{displaymath}
taking a time derivative, using the evolution equations and
integrating by parts gives (note that in the discrete analogue this is
where the ``cross terms'' comes from)
\begin{displaymath}
\frac{d}{dt}{\mathcal E} = \int _0^{\infty} \frac{(v^2-u^2)}{2}|_{x=0}dy -  \int _0^{\infty} (uv)|_{y=0}dx\;.
\end{displaymath}
At the discrete level one is essentially left with the discrete
version of this expression provided operators that satisfy SBP are
used. In particular, the corner's contribution is
\begin{displaymath}
\frac{d}{dt}E =
\ldots + (\Delta y)\frac{1}{2}(v^2-u^2)|_{(0,0)} -  (\Delta x)(uv)|_{(0,0)}\; ,
\end{displaymath}
where the dots represent the contributions from the boundary from
points other than the corner. One has to control the complete corner
term as it will not be canceled or changed by any contribution from
other boundary points. Suppose $\Delta x=\Delta y$. The projector
should set, for example, $\frac{1}{2}(v^2-u^2)|_{(0,0)} -
(uv)|_{(0,0)}$ to zero. Suppose one does not do this but, instead,
controls only the ingoing mode in the $x$ direction (that is, $v$) by
setting it to zero. This does give an energy estimate, since
\begin{displaymath}
\frac{d}{dt}E =
\ldots + \Delta \frac{1}{2}(-u^2)|_{(0,0)} \;, 
\end{displaymath}
which has the appropriate sign (that is, the corner term can be
bounded from above by zero). Suppose on the other hand that one does
not set it to zero but, instead, couples it to the outgoing mode, $v=
s u$, with $s=\pm 1$. Then
\begin{displaymath}
\frac{d}{dt}E = \ldots  -s \Delta (u^2)|_{(0,0)}\;, 
\end{displaymath}
and the energy estimate follows or not, depending on the sign of
$s$.

 
\end{document}